\documentclass[preprint,prd]{revtex4}
\pdfoutput=1
\usepackage{amsfonts,amsmath,amssymb}
\usepackage{enumerate}
\usepackage{hyperref}
\usepackage{graphicx}
\usepackage{subfig}
\newcommand{\be}{\begin{equation}}
\newcommand{\ee}{\end{equation}}

\newcommand{\bea}{\begin{eqnarray}}
\newcommand{\eea}{\end{eqnarray}}

\newcommand{\bal}{\begin{align}}
\newcommand{\eal}{\end{align}}

\renewcommand{\d}{{\rm d}}

\newcommand{\bes}{\begin{subequations}}
\newcommand{\ees}{\end{subequations}}

\newcommand{\nn}{\nonumber}

\newcommand{\ra}{\rightarrow}

\newcommand{\ve}{\varepsilon}

\begin{document}

\title{Time-periodic solutions 
of massive scalar fields \\in AdS background: perturbative constructions}
\author{Nakwoo Kim}
\email{nkim@khu.ac.kr}
\affiliation{Department of Physics
and Research Institute of Basic Science, \\ Kyung Hee University,
Seoul 130-701, Korea }



\begin{abstract}
We consider scalar fields which are coupled to Einstein gravity with a negative
cosmological constant,
and construct periodic solutions perturbatively. In particular, we study
tachyonic scalar fields whose mass is at or above the Breitenlohner-Freedman 
bound in four, five, and seven spacetime dimensions. 
The critical amplitude of the leading order perturbation, for which
the perturbative expansion breaks down, increases as we consider less massive 
fields. We present various examples including a model with 
a self-interacting scalar field which is derived from a consistent truncation of IIB
supergravity. 
\end{abstract}


\maketitle


\section{Introduction}
In the AdS/CFT correspondence \cite{Maldacena:1997re}, one usually relates a strongly
interacting quantum field theory with a classical anti-de Sitter Einstein gravity with a 
negative cosmological constant with matter fields. Replacing a quantum field theory
with a classical equation of motion is certainly a great simplification, but the price
to pay is that one has to go to a higher dimensional spacetime. In broad terms,
the dependence on the {\it radial} direction in the gravity provides the scale 
dependence of physical quantities. A particularly nice property of the AdS/CFT
is that black holes are dual to field theory at finite temperature, so time-dependent
process on the gravity side can in principle describe time evolution of a
thermal system. The quantitative understanding of 
black hole formation within AdS space is thus certainly desirable. 

Recently several groups have studied numerically the formation of a black hole in AdS space with a matter field. A seminal paper along this direction
is \cite{Bizon:2011gg} (see also \cite{Dias:2011ss,Stotyn:2011ns}), where the authors 
presented numerical solutions of the coupled 
nonlinear partial differential equations from Einstein-massless-scalar field
system with a spherically symmetric ansatz. The conclusion drawn from the
data is that AdS spacetime is generically unstable under small perturbations of matter fields,
due to nonlinearity which transfers energy to higher frequency modes. 
However, it was discovered soon that there exist many nonlinearly stable solutions
\cite{Dias:2012tq} and also time-periodic solutions in AdS space 
\cite{Maliborski:2013jca}. The authors of \cite{Maliborski:2013jca} considered
a massless scalar field in AdS${}_5$ space and solved the field equation 
perturbatively and argued for the existence of periodic solutions. Calcellation 
of secular terms through a shift of the frequency is an essential part of the
construction. For related works readers are referred to 
\cite{Fodor:2010hg,Garfinkle:2011hm,Grandclement:2011wz,
Jalmuzna:2011qw,deOliveira:2012dt,Buchel:2013uba,Maliborski:2013via,Fodor:2013lza,
Abajo-Arrastia:2014fma,Maliborski:2014rma,Balasubramanian:2014cja,
Craps:2014vaa,Basu:2014sia,Horowitz:2014hja,Deppe:2014oua,
Dimitrakopoulos:2014ada,Baier:2014ita}.

The aim of this work is to extend the study of time-dependent solutions in
gravity-scalar system to tachyonic fields. In most of the previous works,
probably for definiteness and simplicity, the authors chose massless scalar fields. 
As it is well known however, in AdS space ``massless'' field is not exactly at
the border of stability, which is usually called the Breitenlohner-Freedman (BF)
bound. Stability requirement of a scalar field in AdS${}_{d+1}$ 
for instance is in fact $m^2\ge -\tfrac{d^2}{4\ell^2}$,
where $\ell$ is the curvature radius. According to the AdS/CFT correspondence, 
tachyonic scalars above the BF bound are dual to relevant operators, while 
a massless scalar field is dual to a marginal operator. It is thus an obviously
impending question: whether a tachyonic scalar can also lead to periodic solutions,
and if the answer is yes how much quantitative and qualitative difference they 
have, compared to massless scalars. In the next Section we report the result of
our symbolic computation. For all the tachyonic scalar fields we have considered
we have checked the cancellation of secular terms and explicitly obtained periodic
solutions perturbatively. As it is naturally expected, the radius of convergence 
for the amplitude of perturbation field becomes larger as we consider large
values of $(-m^2)$ values. 
\section{The gravity-scalar system and its perturbative solutions}
Our starting point is the following action of a massive real scalar field field
coupled to Einstein gravity with a cosmological constant $\Lambda$. (We note
that we closely follow the convention of \cite{Bizon:2011gg}.)
\be
S = \int \d^{d+1}x \sqrt{-g} \left( \frac{1}{16\pi G} (R-2\Lambda) 
-\frac{1}{2}(\partial \phi)^2 - \frac{1}{2} m^2 \phi^2 
\right) .
\ee
The spacetime is $d+1$ dimensional, and we consider $\Lambda<0$, i.e. the vacuum 
is anti-de Sitter. We take a spherically symmetric ansatz, and more concretely
the metric is written as
\be
\d s^2 = \frac{\ell^2}{\cos^2 x}
\left( - A e^{-2\delta} \d t^2 + \frac{\d x^2}{A} + \sin^2 x\,  \d\Omega_{d-1}
\right) . 
\ee
Here the metric component fields $A,\delta$, as well as the
matter field $\phi$, depend only on $t,x$. $d\Omega_{d-1}$ denotes the line
element of the $(d-1)$-dimensional unit sphere. The curvature radius $\ell$ is 
determined as $\Lambda=-\tfrac{d(d-1)}{2\ell^2}$.

When one computes the Einstein tensor from the metric ansatz above, at first sight it
looks like there are four non-vanishing and independent components, e.g. $G_{tt}, 
G_{tr}, G_{rr}$ and the components on the sphere $S^{d-1}$. But two of them are 
in fact constraints, which are shown to follow from the remaining equations.
This is of course related to the fact that we have allowed 
non-trivial dependences on two coordinates $t,r$ and their diffeomorphism freedom. 

The scalar equation of motion is given as 
\be
\partial_t (e^{\delta} A^{-1} \partial_t \phi ) - 
\frac{1}{\tan^{d-1}x} \partial_x ( A e^{-\delta} \tan^{d-1} x \partial_x \phi )
+\frac{\Delta(\Delta-d)}{\cos^2 x} e^{-\delta} \phi = 0 \, . 
\label{kge}
\ee
Here we set $m^2=\Delta(\Delta-d)/\ell^2$ and assume that the mass parameter
is above the Breitenlohner-Freedman bound, i.e. $m^2\ge -\tfrac{d^2}{4\ell^2}$. 
Here $\Delta \ge d/2$ is the conformal dimension of the dual operator through AdS/CFT
correspondence.

The two independent equations from the variation of metric are 
\bea
\delta' &=& -  \sin x \cos x ( A^{-2} e^{2\delta} \dot\phi^2 + 
\phi'^2 ) \, , 
\label{ee1}\\
A' &=& A\delta' + \frac{d-2+2\sin^2x}{\sin x\cos x} (1-A) - 
\frac{\Delta(\Delta-d)\sin x}{\cos x} \phi^2 \, . 
\label{ee2}
\eea

We can solve the equations perturbatively around the vacuum AdS solution
$A=1,\delta=0$ 
and $\phi=0$. At first order, we set $\phi=\varepsilon\phi^{(1)}$
for a small parameter $\varepsilon$. If we use the usual technique of separation of
variables $\phi^{(1)} = f(x) \cos \omega t$
the scalar equation \eqref{kge} gives a Sturm-Liouville problem
$L f(x) = \omega^2 f(x)$ with 
\begin{align}
L f(x) &\equiv -\frac{1}{\tan^{d-1}x} \frac{\d}{\d x} \left[
\tan^{d-1} x \frac{\d f}{\d x} \right] + \frac{\Delta(\Delta-d)}{\cos^2 x} f(x) \, . 
\end{align}
It is straightforward to solve this equation. The eigenfunctions and the 
eigenvalues are
\bea
e_j(x) &=& 2 \sqrt{\frac{(j+\Delta/2)\Gamma(j+1)\Gamma(j+\Delta)}{\Gamma(j+d/2)\Gamma(j+\Delta-d/2+1)}} (\cos x)^\Delta P^{d/2-1,\Delta-d/2}_{j}(\cos 2x) 
\, , 
\label{ef}
\\
\omega_j &=& 2j + \Delta \, . 
\eea
Here $P^{a,b}_j(u),\; j=0,1,2,\cdots$ are Jacobi polynomials.
We note that the eigenfunctions are normalized as 
\be
\int^{\pi/2}_{0} e_i(x) e_j(x) \tan^{d-1} x \, \d x = \delta_{ij} \, . 
\ee
At the next order ${\cal O}(\varepsilon^2)$, we can easily
solve \eqref{ee1},\eqref{ee2} and obtain $A=1-\varepsilon^2 A^{(2)}, \delta = \varepsilon^2
\delta^{(2)}$. We choose the convention $\delta(t,x=0)=1-A(t,x=0)=0$ for integration 
constants. More concretely, when we integrate \eqref{ee1} 
\be
\delta^{(2)}(t,x) = - \int^x_0 \sin y \cos y \left( (\partial_t \phi^{(1)}(t,y))^2 + (\partial_y\phi^{(1)}(t,y))^2 \right) \d y . 
\label{d2}
\ee
Similarly we get 
\bea
A^{(2)}(t,x) &=& \frac{\cos^d x}{\sin^{d-2}x} 
\int^x_0 \tan^{d-1} y
\Big[
(\partial_t \phi^{(1)}(t,y))^2 + (\partial_y\phi^{(1)}(t,y))^2 
\nn\\ 
&&  + \frac{\Delta(d-\Delta)}{\sin y
\cos y} (\phi^{(1)}(t,y))^2 
\Big]
\, \d y \, . 
\label{a2}
\eea

In the next order ${\cal O}(\varepsilon^3)$ we need to solve the scalar 
equation which now becomes an in-homogeneous second order differential 
equation.
\bea
(\partial_t^2 + L) \phi^{(3)}(t,x) &=& 
\frac{\Delta(\Delta-d)}{\cos^2 x} \delta^{(2)} \phi^{(1)} 
-\partial_t \left[ (\delta^{(2)}+A^{(2)}) \partial_t \phi^{(1)} \right]
\nn\\
&& 
-\frac{1}{\tan^{d-1} x} \partial_x \left[
(\delta^{(2)}+A^{(2)}) \partial_x \phi^{(1)} \right]
\, . 
\label{p3}
\eea
Here the point is that on the right hand side of the above equation there appears
a product of three harmonic functions like $(\cos \omega t)^3$. Using the 
elementary algebra of trigonometric functions, it gives rise to 
{\it secular} modes whose frequency is the same as one of 
the original frequencies  $\omega_j=\Delta+2j$.
Naively this means that the amplitude
of the resonant modes increases linearly with time, but as it is well known this kind
of instability is unphysical if it can be absorbed by shifting the frequency 
$\omega \ra \omega + \varepsilon^2 \omega^{(2)}$. It has been verified in 
\cite{Maliborski:2013jca} that, for $d=4$ (AdS${}_5$) and a massless scalar field, 
if we start with a single mode at ${\cal O}(\varepsilon)$
the secular terms are cancelled perturbatively up to fairly high orders in 
$\varepsilon$. 
For AdS${}_5$ and the 
the lowest lying mode $j=0$, the frequency as 
a function of the perturbative parameter $\varepsilon$ is found as 
\be
\Omega = 4+ \frac{464}{7} \ve^2 + \frac{45614896}{11319} \ve^4 + \cdots \, . 
\label{ads5res}
\ee
In \cite{Maliborski:2013jca} it is reported that the coefficients were obtained 
up to $\ve^{16}$. Through the Pad\'e approximation the series seems to be
convergent with radius of convergence $\ve \approx 0.09$.

From the analytic expression of the perturbative solution, we may extract
a lot of data which can help us understand the time-evolution of our solution. 
Let us take the function $A$ for example.
It is obvious that $A=0$ at a particular point in the spacetime
implies the formation of a black hole. From the expression for $A$ which is exact
up to the order of ${\cal O}(\ve^{20})$ we have created plots for the time-oscillation 
for different values of $\ve$. The minimum of $A$ decreases for larger $\ve$, 
and if we extrapolate our perturbative solution to bigger values of $\ve$, 
$A$ hits zero at $\ve\approx 0.11$. 

\begin{figure}[h]
\begin{tabular}{c  c}
\subfloat[Oscillation of $A(x)$, for $\ve=0.06$] {\includegraphics[scale=0.35]{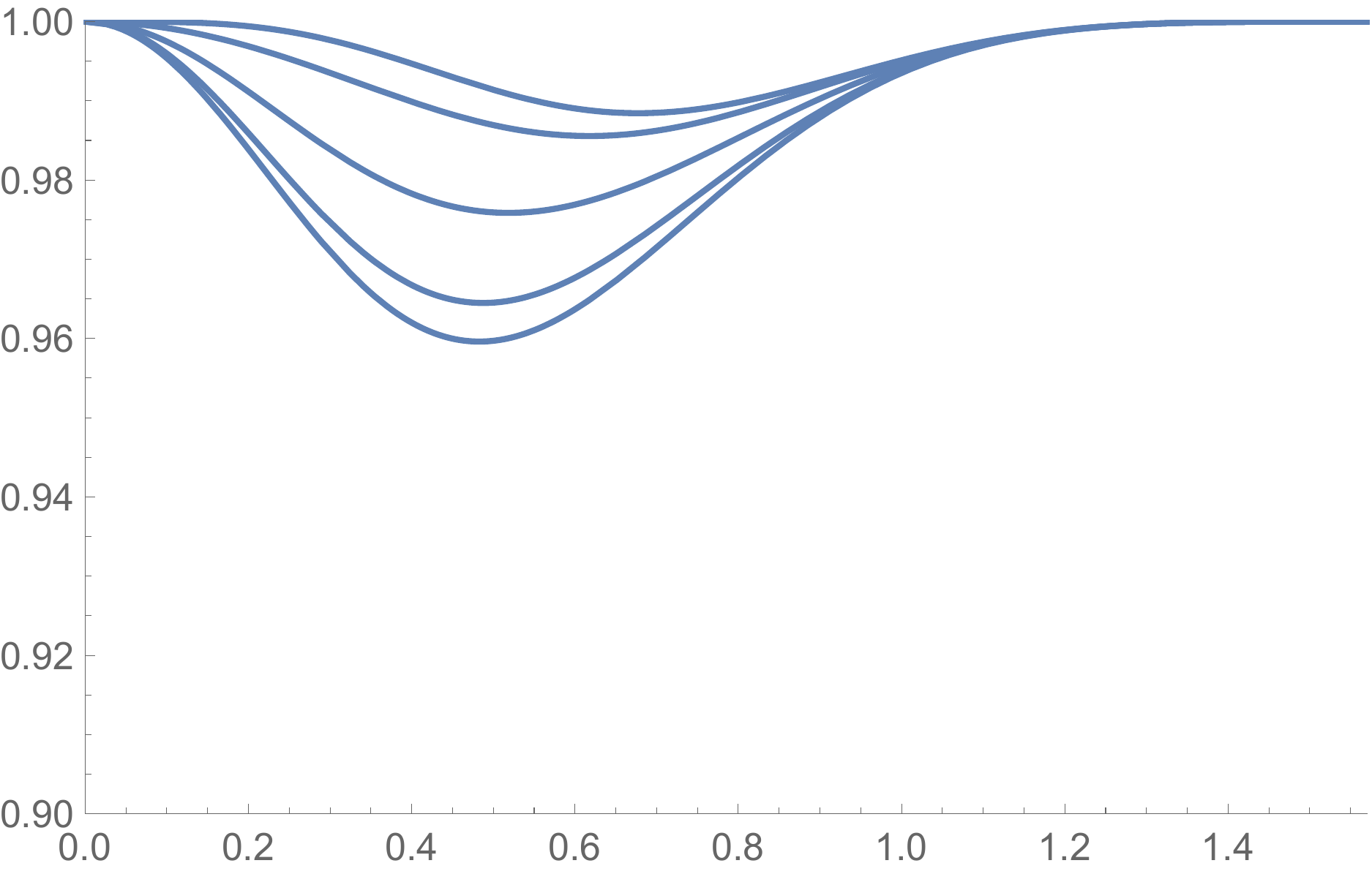}} &  
\subfloat[min$(A)$ vs. $\ve$]{\includegraphics[scale=0.35]{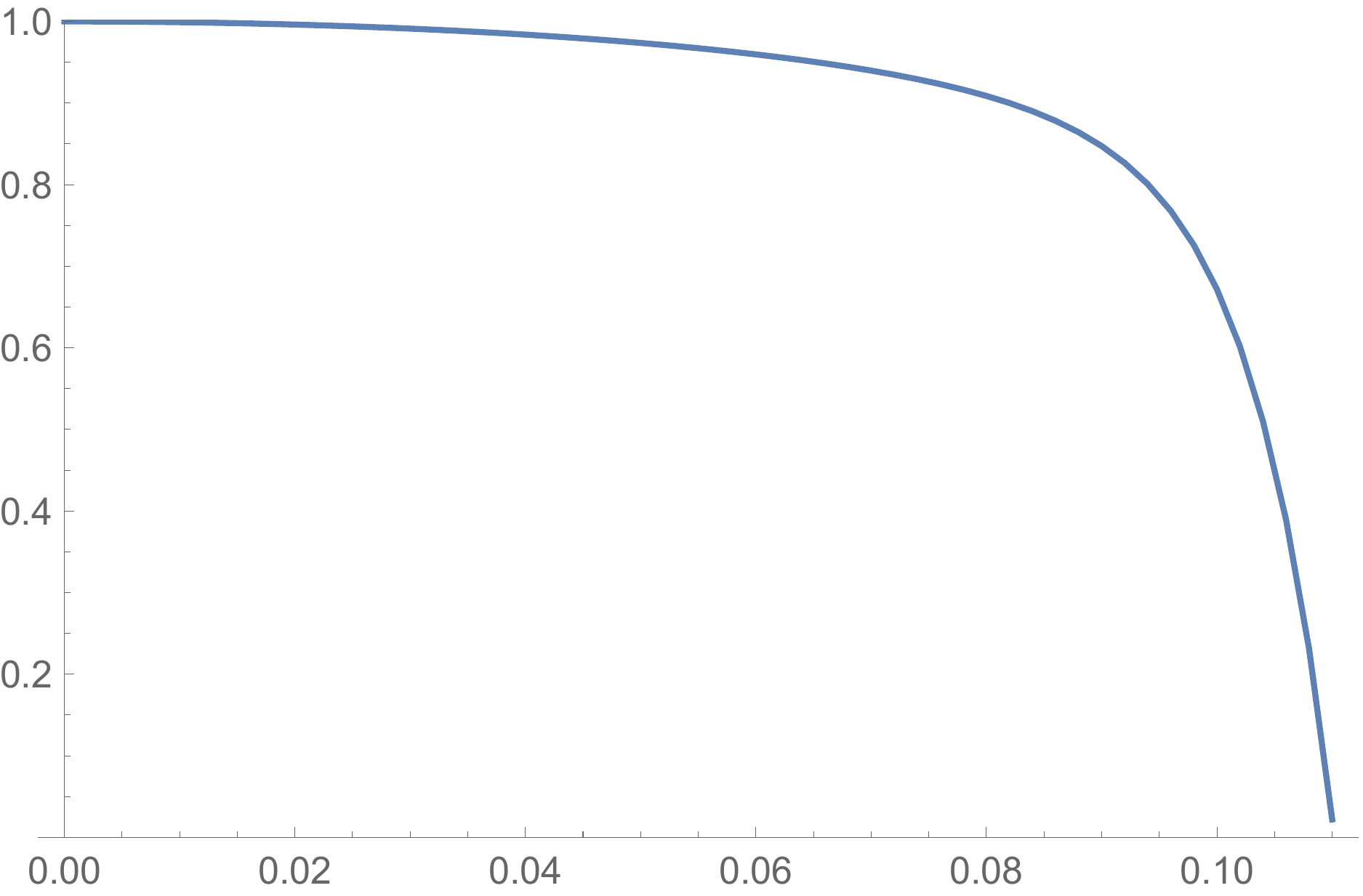}}
\end{tabular}
\caption{The plot in the left panel shows the oscillation of $A(t,x)$. The minimum value of $A$ decreases for larger $\ve$. The plot 
in the right panel shows min$(A)$ as a function of $\ve$.}
\label{plot1}
\end{figure}
\subsection{Massive scalars in AdS${}_5$}
We have written a code which constructs periodic solutions perturbatively in {\it Mathematica} 
and have 
confirmed the result for the case of a massless scalar field in AdS${}_5$ agrees with 
\cite{Maliborski:2013jca}. In fact
we pushed the computation to ${\cal O}(\ve^{20})$: the coefficients of $\ve^{18},\ve^{20}$ in \eqref{ads5res} are approximately
$3.92591\times 10^{17}, 4.45447\times 10^{19}$.  
The Pad\'e approximation at $(10,10)$ then 
gives the pole of the denominator at $\ve=0.0904562$.
 On our laptop with an 
2.7GHz Intel i7 CPU and 16 GB RAM, the last step of calculating this coefficient
at ${\cal O}(\ve^{20})$ 
took less than 3 hours and 10 minutes. 

The first choice of our own 
is a scalar field exactly at the BF bound ($\Delta=2$, or equivalently $m^2=-4/\ell^2$). 
For the lowest lying mode 
the probe limit gives eigenfrequency $\omega=2$. Our perturbative algorithm gives
\bea
\Omega &=& 2+\frac{62}{15} \varepsilon^2+\frac{31373}{2250} \varepsilon^4+\frac{1757780088437
   }{24169635000}\varepsilon^6+\frac{537359617120101825761}{1278264665452500000} \varepsilon^8
   \nn\\
   &+& 
 \frac{1268572361264125960914631343583143413}{483709750330891209012418781250000} \varepsilon^{10}
 \nn\\
 &+&
 \frac{1459283228526801137059175769554860613450261076853469483
  }{85131519913245630307027620371897681405325000000000}
   \varepsilon^{12}
+\cdots
\eea
We have obtained the coefficients of up to $\ve^{20}$. They all turn out to 
be rational numbers whose numerator and denominator have too many digits
to be explicitly reported here. The coefficients of $\ve^{14},\ve^{16},\ve^{18},
\ve^{20}$ are approximately $1.1582059\times 10^5, \, 8.0214964\times   10^5, \, 
5.6628666\times 10^6, \, 4.0593615\times10^7$.

Based on this result we also performed the Pad\'e approximation. From $(4,4)$ up
to $(10,10)$, the (smallest) zero of the denominator is respectively $
0.411687, \, 
0.380910,\, 0.368009,\, 0.360290$.

We have repeated a similar computation for $\Delta=3$ or $m^2=-3/\ell^2$. 
\bea
\Omega&=&
3+\frac{297 }{14}\varepsilon^2+\frac{11388681}{27440}\varepsilon^4+
\frac{75814410351189977829 }{6895049537868800}\varepsilon^6
\nn\\
&+&
\frac{440953730050912073171536929147}{1332903538719975878656000
}\varepsilon^8
\nn\\
&+&
\frac{108962184535866721154985183970785991244847410023180150827 
}{10174886986824762523197846218808661282081341440000}
\varepsilon^{10}+\cdots
\eea
For the next coefficients our result gives for the coefficients of 
$\ve^{12},\ve^{14},\ve^{16},\ve^{18},\ve^{20}$ approximate
values $3.6360632\times
   10^8,1.2769397\times
   10^{10},4.5987515\times
   10^{11},1.6888399\times
   10^{13},6.2996028\times
   10^{14}$.
The Pad\'e approximation gives that the upper bound for the perturbative 
approach to be well-behaved is $\ve\approx 0.157957$.

An interesting variation of this system is given by a truncation of
supergravity, from the solutions of the form $AdS_5\times X^5$ in IIB
supergravity where $X^5$ is a Sasaki-Einstein manifold \cite{Gauntlett:2009zw}. 
In particular, a further
truncated theory with a vector field and two real scalars with a nontrivial
potential function has been used to address holographic superconductors
\cite{Gubser:2009qm}.  For our purpose we turn off the vector field as well
as the axion. Then the scalar potential adjusted to our convention is written as 
\be
V(\phi) = - \frac{1}{2\ell^2} \left( -4 + \cosh^2 \frac{\sqrt{6}\phi}{2} \left(
5-\cosh\sqrt{6}\phi
\right)
\right)
\ee
In the small field limit the mass of the scalar corresponds to $\Delta=3$. 
We have confirmed that the cancellation of secular terms persist also in this
supergravity-inspired model. 
\bea
\Omega&=&3+\frac{837}{35} \varepsilon^2+\frac{18123993
  }{42875} \varepsilon^4+\frac{1022167072159904258901 }
   {102073404519520000}\varepsilon^6
   \nn\\
   &+&\frac{1885826584327612453573347913521573 
  }{6924534314390082309920000000} \varepsilon^8
   \nn\\
   &+&\frac{4854622063875589224275650949019735931691812729968299301296434317 }{606704146497857938746558919904014257651677884288000000000}
   \varepsilon^{10}
   \nn\\
   &+& \cdots
\eea
We have also obtained more coefficients up to $\ve^{20}$: they are 
$2.4695488\times 10^8,\, 7.8888024\times 10^9, 2.5852724\times 10^{11},
8.6415084\times 10^{12},2.9344207\times 10^{14}$.
The Pad\'e approximation at $(10,10)$ gives that the radius of convergence for $\ve$ is 
$0.165696$ and there is no huge difference from the previous example of $\Delta=3$.

\subsection{Massless and massive scalars in $AdS_7$}
We can repeat the same analysis for $d=6$ case. Again at any order of the 
perturbative computation the fields are expressed as a finite order polynomial
of $u=\cos x$. For a massless scalar field, in the probe limit the eigen-frequency
is $\omega=6$. Explicit computation gives 
\bea
\Omega &=& 
6 + \frac{133920}{143}\ve^2+\frac{204857013644928}{347980633}\ve^4
\nn\\
&&
 + \frac{6653917224931500928527205438989898}{13966872826194750738453177} \ve^6
\nn\\
&&      +\frac{1080461333185973832680808465193246135959458551807301}{2505052939963175783424913698557925050655}\ve^8+ \cdots
\eea
We obtained the coefficients up to $\ve^{20}$. The coefficients of $\ve^{10},\cdots,
\ve^{20}$ are $4.16417\times 10^{14},\, 4.1925\times
   10^{17},\, 4.34786\times
   10^{20},\, 4.60964\times
   10^{23},\, 4.97161\times
   10^{26},\, 5.43594\times 10^{29}$.
The Pad\'e approximation at $(n,n)$ for $n=2,\cdots,10$ gives the poles of the
denominator at
$
0.032326,0.0304412,0.029661,0.0292566$. When compared to 
the case of AdS${}_5$, the coefficients are larger and the pole of Pad\'e approximant
is smaller. This means that the perturbative expansion breaks down more easily 
for small amplitude of $\phi^{(1)}$. Another way to see this is to check 
how many modes are turned on for a specific order of $\ve$. At $\ve^{20}$, 
the scalar field includes $e_{70}$. On the other hand, for $AdS_5$ the highest
mode at $\ve^{20}$ is $e_{50}$.

We have repeated the computation for a tachyonic scalar field with $\Delta=3,4,5$.
Firstly for $\Delta=3$, or $m^2=-9/\ell^2$.
\bea
\Omega &=& 3 
+
\frac{3807}{280}\ve^2+\frac{2704629609}{21952000}\ve^4+
\frac{22814710893326488039461}{14774928824320000000}\ve^6
\nn\\
&& +\frac{11684631773098620212295629421580959}{544768173672961638400000000000}\ve^8
+\cdots
\eea
And the next coefficients for $\ve^{10},\cdots, \ve^{20}$ are 
$
3.16957\times 10^5,\, 4.87182\times 10^6,\, 7.69766\times
   10^7,\, 1.24135\times 10^9,\, 2.03364\times
   10^{10},\, 3.37368\times 10^{11}
$.
The Pad\'e approximation at $(n,n)$ for $n=4,6,8,10$ exhibit a pole at 
$0.264111,\, 0.248389, \, 0.242002, \, 0.238681$.

Secondly for $\Delta=4$, or $m^2=-8/\ell^2$. The frequency is given as 
\bea
\Omega &=& 4 
+ 
\frac{3152}{35}\ve^2+\frac{24139995472}{4244625}\ve^4+
\frac{89200146157625691820278256}{190178211481119736875}\ve^6
\nn\\
&& +
\frac{961459118
   126637937051446867780955648086736}{22263
   941138510438405094532804453125}\ve^8 
+\cdots .
\eea
The next coefficients for $\ve^{10},\cdots, \ve^{20}$ are 
$4.23236\times 10^9,\,4.31887\times
   10^{11},\,4.53397\times
   10^{13},\,4.86117\times 10^{15},\, 5.29766\times
   10^{17},\, 5.84892\times 10^{19}$.
 The Pad\'e approximation at $(n,n)$ for $n=4,6,8,10$ exhibit a pole at 
 $0.101701, \, 0.0958928, \, 0.0935393, \, 0.0923499$.
 
Finally for $\Delta=5$, or $m^2=-5/\ell^2$. The frequency is given as 
\bea
\Omega &=& 5
+ 
\frac{103375}{308}\ve^2+\frac{1065400702671875}{13674076416}\ve^4
\nn\\
&&
+\frac{867158669318199310085443515535234375}{37159737704925947186764136448}\ve^6
\nn\\
&& +\frac{2031755203183353658899088230995968728695683830185546875}{260115931529596661308343233550649096051621888}\ve^8
+\cdots . 
\eea
And the next coefficients for $\ve^{10},\cdots, \ve^{20}$ are 
$2.78552\times 10^{12},\, 1.03514\times
   10^{15},\, 3.96003\times
   10^{17},\, 1.54806\times
   10^{20},\, 6.15393\times
   10^{22},\, 2.4793\times 10^{25}$.
 And the Pad\'e approximation at $(n,n)$ for $n=4,6,8,10$ exhibit a pole at 
 $0.0532139,\, 0.0501531,\, 0.0488935, \, 0.0482435$.
\subsection{Massless scalar in AdS${}_4$}
In this subsection we address the case of an odd $d$, 
in particular AdS${}_4$. The general analysis here
is rather cumbersome, because from the next order in perturbation at 
${\cal O}(\ve^2)$ the perturbative fields are not 
given as  a polynomial in $u=\cos x$. This means that we have to 
deal with a summation over all the eigenmodes. However, 
one can show that at ${\cal O}(\ve^3)$ the secular modes can be removed through 
a shift of the frequency in the scalar equation, just like previous examples. 

More concretely let us consider perturbation with a massless scalar field which has
the smallest frequency, $\omega=3$.
\be
\phi^{(1)} = \ve e_0 (u) = \ve \sqrt{\frac{32}{\pi}} u^3 \cos (3t) \, . 
\ee
Then from \eqref{d2} we obtain 
\be
\delta^{(2)} = \frac{12}{\pi} \left(
2(u^6-1) - (3u^8-2u^6-1) \cos (6t)
\right) \, , 
\ee
which can be expressed as a linear combination of eigenmodes $e_j(u)$.
For the function $A$ however, we obtain as a function of $u=\cos x$ given 
as follows:
\be
A^{(2)} = \frac{6u^3}{\pi}
\left(
\frac{3  \cos^{-1} u }{\sqrt{1-u^2}}
+ u \left(
3-6u^2 + 8u^2(u^2-1) \cos (6t)
\right)
\right) \, . 
\ee

This obviously involves an infinite sum over the eigenmodes $e_j(u)$
in \eqref{ef}, which are all polynomials in $u$. The expansion coefficients
for the right hand side of \eqref{p3} as $\sum_{j=0}^{\infty} f_j(t) e_j(u)$
can be worked out and the result is
\bea
f_0(t) &=& -\frac{459(8\cos 3t - 5 \cos 9t)}{16\pi}\, ,
\\
f_1(t) &=& \frac{9\sqrt{3}(1374 \cos 3t - 595 \cos 9t )}{160\pi}\, ,
\\
f_2(t) &=& \frac{9\sqrt{6}(59\cos 3t - 140 \cos 9t)}{160\pi}\, ,
\\
f_3(t) &=& \frac{14607\sqrt{10} \cos 3t}{5600\pi}\, ,
\\
f_4(t) &=& \frac{9\sqrt{15}(202\cos 3t + 175 \cos 9t)}{5600\pi} \, ,
\\
f_j(t) &=& \frac{162\sqrt{2}}{\pi} \cdot\frac{ (-1)^j (2j+3)(j^2+3j+8)\cos 3t}
{j(j-1)(j+3)(j+4)(j+1)^{3/2}(j+2)^{3/2}}
,  \quad  j\ge 5 \, .  
\eea
The appearance of
$\cos 3t$ and $\cos 9t$ is easy to understand, 
since at ${\cal O}(\ve^3$) we are dealing with
$\cos^3 3t$. If we recall that the eigenfrequency for $e_j$ is $\omega_j=3+2j$, 
potentially there can be resonances for $\omega_0=3$ and $\omega_3=9$.  But
as we see in the above, $f_3$ does {\it not} contain $\cos 9t$: this rather
miraculous cancellation of secular terms applies to all other examples 
discussed in \cite{Maliborski:2013jca}\cite{Craps:2014vaa} and
 this paper so far. $f_0$ contains a resonance term, but we can cancel it 
 through renormalization of the frequency $\omega\ra \Omega=\omega+ \ve^2 \omega^{(2)}$,
 with
\be
\Omega = 3 + \frac{153}{4\pi} \ve^2\, . 
\ee
Integration of \eqref{p3} is now straightforward using the technique of
separation of variables. 
It will be interesting to compute higher order terms in $\Omega$, but we will
leave it for a future work. 

\section{discussion}
We have so far analysed the perturbative computation of classical scalar-Einstein
gravity equations by extending previous works on massless scalars to the case of 
tachyonic ones. This work stands also as a technical improvement, since we have 
pushed 
the perturbative expansion to ${\cal O}(\ve^{20})$, while Ref.\cite{Maliborski:2013jca}
reported results upto ${\cal O}(\ve^{16})$. 
Our result confirms that the periodic solutions and the associated 
removal of secular terms in \cite{Maliborski:2013jca} persist for massive
scalars. The central quantitative result of ours is the change of frequency 
renormalization as a function of the mass of the scalar fields. We confirmed
the natural prediction that the perturbative series should be valid for larger
amplitudes as we decrease $m^2$.  

It will be interesting if one can generalize
our analysis to non-integer values of $\Delta$, but in that case - just like a 
massless scalar in AdS${}_4$ - higher order 
configurations in general cannot be expressed as a finite sum over the normal 
modes of the probe scalar equations so it will be
difficult to automatize the computation. It will be very nice if we can find
the exact mass dependence of the radius of 
convergence for the perturbation parameter $\ve$, for general $\Delta$ and $d$. 
It will be also interesting to study different matter fields or modified gravity 
theories, for instance Gauss-Bonnet theory which through AdS/CFT correspondence
corresponds to $1/N$ corrections on the dual field theory side.


\begin{acknowledgments}
We are grateful to L.A. Pando Zayas for drawing our attention to \cite{deOliveira:2012dt}, and also for comments and discussions.
This work was supported by the sabbatical leave petition program (2012) of 
Kyung Hee University (KHU-20120649), National Research Foundation of Korea 
(NRF) grants funded by the Korea government (MEST) with grant No. 2010-0023121 
and  No. 2012046278.
\end{acknowledgments}
\bibliography{pp}
\end{document}